\documentclass[twocolumn,aps,prl,groupedaddress]{revtex4-1}

\usepackage{graphicx}
\usepackage{dcolumn}
\usepackage{bm}
\usepackage{amsmath,amssymb,amsfonts,bbm}
\usepackage{color}
\usepackage{braket}

\begin{document}

\title{Magnus Hall Effect}
\author{Micha{\l} Papaj}
\author{Liang Fu}
\affiliation{Department of Physics, Massachusetts Institute of Technology, Cambridge, Massachusetts 02139, USA}

\begin{abstract}
A new type of linear response Hall effect is predicted in time-reversal-invariant systems with built-in electric field at zero magnetic field. The Hall response results from a quantum Magnus effect where a self-rotating Bloch electron wavepacket moving under electric field develops an anomalous velocity in the transverse direction. We show that in the ballistic limit the Magnus Hall conductance measures the distribution of Berry curvature on the Fermi surface.
\end{abstract}

%\pacs{}

\maketitle

\textit{Introduction}.--- Studies of various Hall effects led to significant progress throughout the history of solid state physics. Starting with the classical Hall effect\cite{HallNewActionMagnet1879}, anomalous Hall effect \cite{NagaosaAnomalousHalleffect2010}, spin Hall effect \cite{SinovaSpinHalleffects2015},  thermal Hall effect \cite{BanerjeeObservationhalfintegerthermal2018, KasaharaMajoranaquantizationhalfinteger2018}, quantum Hall effect \cite{KlitzingNewMethodHighAccuracy1980a}, quantum spin Hall effect \cite{KaneQuantumSpinHall2005a} and quantum anomalous Hall effect \cite{HaldaneModelQuantumHall1988c, ChangExperimentalObservationQuantum2013a, LiuQuantumAnomalousHall2016} have been discovered.
Among these, classical and anomalous Hall effects appear in time-reversal-breaking systems, where an applied electric field induces a transverse charge current.

An intrinsic contribution to anomalous Hall effect is associated with Berry curvature, a fundamental ingredient of modern band theory derived from electron's wavefunction within the unit cell \cite{KarplusHallEffectFerromagnetics1954}. When a Bloch electron is accelerated under an electric field,  the redistribution of electron density within the unit cell changes with increasing momentum, thus giving rise to an anomalous velocity proportional to Berry curvature. In time reversal breaking systems, total Berry curvature of occupied states can be nonzero, resulting in intrinsic anomalous Hall effect. The impact of Berry curvature on transport phenomena has attracted tremendous interest \cite{XiaoBerryphaseeffects2010, MooreConfinementInducedBerryPhase2010, MorimotoSemiclassicaltheorynonlinear2016, MorimotoTopologicalnaturenonlinear2016, ChaudharyBerryelectrodynamicsAnomalous2018, RudnerBerryogenesisselfinducedBerry2018}. 

On the other hand, there exists a large class of time-reversal-invariant, inversion-breaking materials which feature large Berry curvature $\bm \Omega(\bm k)$ in momentum space, especially near the gap edge or band crossing points. Examples include two-dimensional transition metal dichalcogenides (TMDs) \cite{QianQuantumspinHall2014}, graphene multilayers \cite{McCannelectronicpropertiesbilayer2013, RozhkovElectronicpropertiesgraphenebased2016} and heterostructures \cite{YankowitzEmergencesuperlatticeDirac2012}, topological insulator surface states \cite{FuHexagonalWarpingEffects2009a} and Weyl semimetals \cite{HasanDiscoveryWeylFermion2017, ArmitageWeylDiracsemimetals2018, YanTopologicalMaterialsWeyl2017}. Due to time reversal symmetry, the distribution of Berry curvature satisfies $\bm \Omega(\bm k)= -\bm \Omega(-\bm k)$. It is an intriguing question whether such a distribution of Berry curvature with zero total can lead to any interesting phenomena in charge transport.

\begin{figure}[b]
\includegraphics[width=0.49\textwidth]{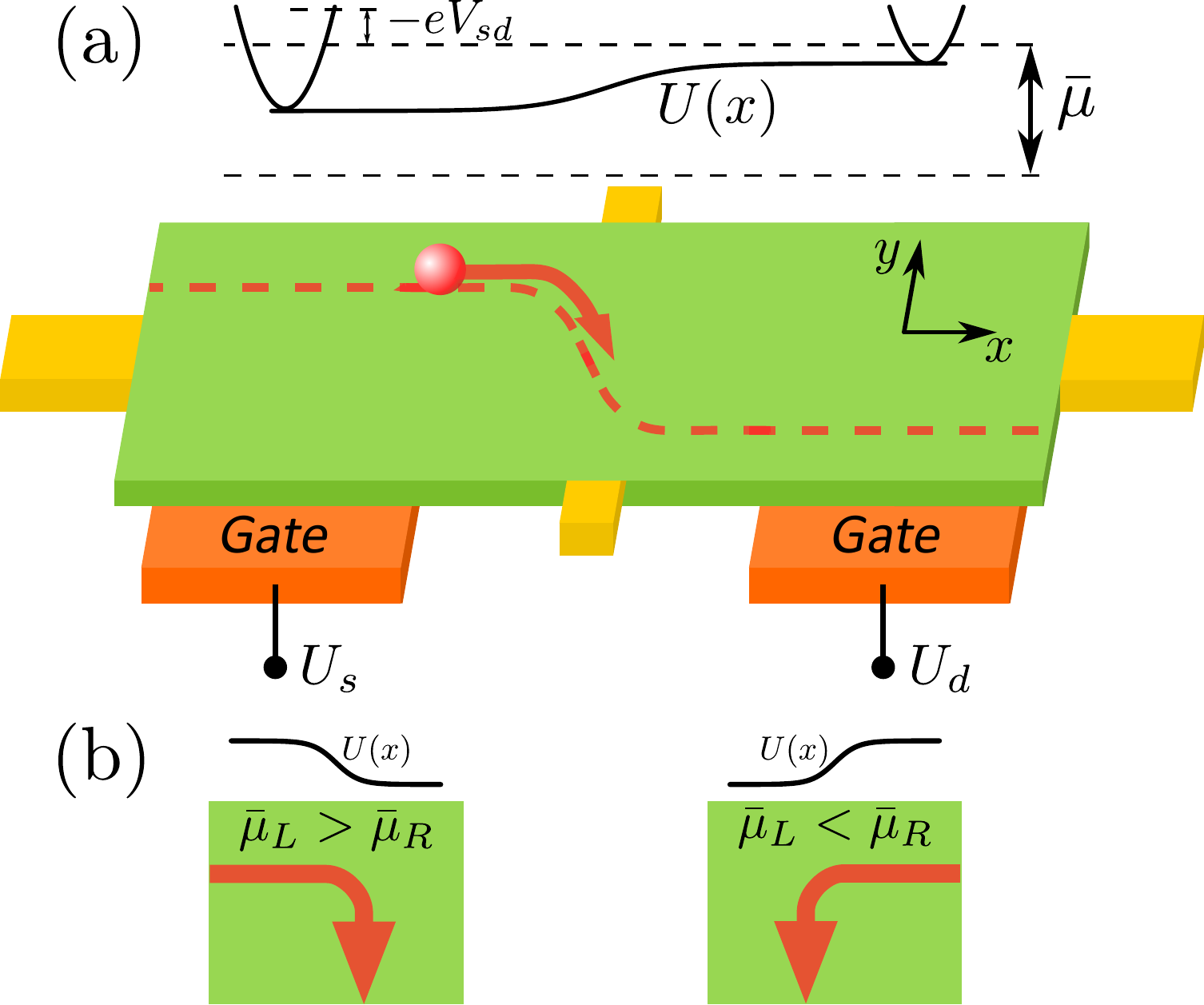}
\caption{\label{fig:device}(Color online) (a) Geometry of the device for Magnus Hall effect. Source and drain regions are separated by a segment with built-in electric field. Electron that exits the source exhibits a Magnus shift $\Delta y_A$. Potential energy $U(x)$ profile in the central region determines the position of the band bottom. (b) Direction of Hall current is preserved under inversion of both the direction of potential drop and the electrochemical potential bias.}
\end{figure}

In this work, we demonstrate a new type of linear-response Hall effect induced by Berry curvature and built-in electric field in mesoscopic systems under time-reversal-symmetric condition. We consider electron transport in a Hall bar device made of a 2D material, such as bilayer graphene or transition metal dichalcogenide. In our setup, source and drain regions have different carrier densities, which can be achieved by local bottom gates. The difference in electrostatic potential energy due to bottom gates $U_s - U_d \equiv \Delta U$ is accompanied by a built-in electric field in the junction at equilibrium, as shown in Fig. \ref{fig:device}(a). We study electrical current in linear response to applied bias voltage $V_{sd}$. Since electrons moving from source to drain have nonzero net velocity, their wave packets can carry orbital angular momentum and nonzero net Berry curvature. The motion of chiral Bloch electrons under built-in electric field leads to a quantum analog of Magnus effect: as an electron traverses the junction, its center of wavepacket acquires a transverse shift. This in turn gives rise to a transverse current linearly proportional to the bias voltage, i.e., a Hall effect. We term this phenomenon ``Magnus Hall effect''. It occurs  in nonmagnetic systems at zero magnetic field, but relies on the built-in electric field. We show that in the ballistic limit, the Hall conductance is directly related to the Berry curvature distribution of the underlying material.

Magnus Hall effect is intimately related to nonlinear Hall effect, where recent theory predicted that in time-reversal-invariant materials the dipole moment of Berry curvature, i.e.,  Berry curvature dipole, can induce a nonlinear Hall effect, where the transverse current depends quadratically on the applied electric field \cite{SodemannQuantumNonlinearHall2015}.  This phenomenon has been subsequently observed in bilayer WTe$_2$ \cite{MaObservationnonlinearHall2019, KangNonlinearanomalousHall2019}. In the case of nonlinear Hall effect, reversing the electric field doesn't change the transverse current. In Magnus Hall effect, transverse current is preserved under reversing both source-drain voltage and the direction of built-in electric field as presented in Fig.\ref{fig:device}(b). In an alternative scenario where the electric field arises from the source-drain voltage $V_{sd}$ itself instead of bottom gates, the transverse current becomes second order in $V_{sd}$, resulting in nonlinear Hall effect.
Our work thus opens a pathway to ``Hall diodes'' for high-frequency nonlinear transport based on quantum materials that have a significant distribution of Berry curvature.

\textit{Electronic Magnus effect}.---
First we shall consider an electron wave-packet which travels inside a sample. The sample has a short segment (between $x=0$ and $x=L$) within which there is spatially varying potential energy $U(x)$ arising from different gate voltages on opposite sides (this corresponds to built-in electric field $\mathbf{E} = \frac{1}{e} \frac{\partial U}{\partial x} \hat{x} = E_x (x) \hat{x}$). Inside this region, motion of the wave-packet is described by the semiclassical equations of motion \cite{XiaoBerryphaseeffects2010}:

\begin{equation}
\label{eq:semiclassical}
\dot{\mathbf{r}} = \frac{1}{\hbar} \frac{\partial \epsilon_\mathbf{k}}{\partial \mathbf{k}} - \frac{1}{\hbar} \mathbf{\Omega} \times \frac{\partial U}{\partial \mathbf{r}}, \quad \dot{\mathbf{k}} = - \frac{1}{\hbar} \frac{\partial U}{\partial \mathbf{r}}
\end{equation}

Since the built-in electric field is small, wave packet momentum $\mathbf{k}$ does not change substantially under acceleration. Therefore, the transit time through the electric field region of an incident electron with velocity $(v_x, v_y)$ is simply $t = L/v_x$.  During this time, the electron with $v_y\neq 0$ will also travel in the $y$ direction. Importantly, between $x=0$ and $x=L$ there will be an additional displacement of wave packet along $y$ due to anomalous velocity. This displacement is given by:
\begin{align}
\Delta y_A &= - \int_0^t \frac{\Omega(\mathbf{k})}{\hbar} \frac{\partial U}{\partial x}  dt' \approx -\frac{1}{\hbar v_x} \Omega(\mathbf{k}_0) \int_0^L \frac{\partial U}{\partial x}  dx = \notag \\ &= \frac{1}{\hbar v_x} \Omega(\mathbf{k}_0) \Delta U
\end{align}
with $\Delta U = - \int_0^L \frac{\partial U}{\partial x} dx$ being the difference in potential energy. Therefore, an electron wave packet with non-zero Berry curvature moving through the region of electric field will acquire an additional shift in the direction perpendicular to the electric field as schematically shown in Fig. \ref{fig:device}(a). This in turn leads to a current density in $y$ direction by integrating single wave-packet contribution $-e \Delta y_A/t = - e/(\hbar L) \Omega(\mathbf{k}_0) \Delta U$ over occupied states. This transverse current vanishes in equilibrium. However, in the current carrying state, the modes with positive and negative velocity are not equally occupied. Since in systems with time-reversal symmetry Berry curvature is opposite for $\mathbf{k}$ and $-\mathbf{k}$ states (which also have opposite velocities), this effect can lead to a Hall current even in nonmagnetic materials. To see more clearly how the Hall current arises here, we employ Boltzmann transport equation.

\textit{Boltzmann transport equation solution}.---
To describe mesoscopic electron transport in a 2D system we use a collisionless Boltzmann equation (BE):
\begin{equation}
\label{eq:boltzmann_eq}
\frac{\partial f}{\partial t} + \dot{\mathbf{r}} \cdot \frac{\partial f}{\partial \mathbf{r}} + \dot{\mathbf{k}} \cdot \frac{\partial f}{\partial \mathbf{k}} = 0
\end{equation}
where $f(\mathbf{k}, \mathbf{r})$ is the occupation distribution function. Later we shall extend the discussion to non-ballistic case with finite relaxation time. We are looking for a stationary state distributions, so $\frac{\partial f}{\partial t} = 0$. We solve the Boltzmann equation in a geometry of a stripe of infinite width in $y$ direction and finite length $L$ in $x$ direction. At $x<0$ and $x>L$ we have a source and a drain. Since the system is translationally invariant in $y$ direction, the stationary distribution function will be independent of $y$.

The wave packets evolve according to semiclassical equations given by Eq.\eqref{eq:semiclassical}. The energy of the electrons in the segment is:
\begin{equation}
\label{eq:spatial_energy}
\epsilon(\mathbf{k}, \mathbf{r}) = \epsilon_\mathbf{k} + U(x)
\end{equation}
where $\epsilon_\mathbf{k}$ is the band energy. In equilibrium, the solution is given by Fermi-Dirac distribution with constant electrochemical potential $\bar{\mu}$, but with spatially changing energy \eqref{eq:spatial_energy} and can be expressed as:
\begin{equation}
f_{0}(\mathbf{k}, \mathbf{r}) = \left(e^{\beta(\epsilon(\mathbf{k}, \mathbf{r}) - \bar{\mu})} + 1 \right)^{-1}
\end{equation}
$f_{0}(\mathbf{k}, \mathbf{r})$ is a solution of Boltzmann equation \eqref{eq:boltzmann_eq} as can be verified using Eqs.\eqref{eq:semiclassical}. This solution guarantees that no current is flowing in the system as at each energy and position the number of states with $\mathbf{k}$ and $-\mathbf{k}$ is equal.

To obtain a solution that corresponds to a steady current flow, we apply a small bias $V_{sd}$ in a form of imbalance of electrochemical potentials between source and drain, so that $\bar{\mu}_D = \bar{\mu}$ and $\bar{\mu}_S = \bar{\mu} - eV_{sd} = \bar{\mu} + \Delta \bar{\mu}$. We have to solve BE with boundary conditions that take into account the presence of reservoirs at $x=0$ and $x=L$. Taking into consideration the device geometry of our system, we now look for a solution in lowest order of the perturbation in the electrochemical potential imbalance. To achieve this, we write $f = f_0 + f_1$, where $f_1$ is the nonequilibrium part first order in the perturbation $\Delta \bar{\mu}$. This gives an equation for $f_1$:

\begin{equation}
\frac{1}{\hbar} \frac{\partial f_1}{\partial x} \frac{\partial \epsilon_k}{\partial k_x}  - \frac{1}{\hbar} \frac{\partial U}{\partial x} \frac{\partial f_1}{\partial k_x} = 0
\end{equation}

Since we assumed that $U(x)$ is slowly varying, the second term on the left hand side is small, so we drop it and we arrive at $\frac{\partial f_1}{\partial x} = 0$. Therefore, $f_1(\mathbf{k},\mathbf{r})$ only depends on $\mathbf{k}$. We can now determine its form from the boundary condition. The larger electrochemical potential of the source region results in a surplus of electrons entering the system at $x=0$ interface with positive $v_x$ velocity and propagating across the device without any scattering in the ballistic limit. This boundary condition gives us:

\begin{align}
f_1(\mathbf{k},\mathbf{r}) =
\begin{cases}
\left(-\frac{\partial f_0 }{\partial \epsilon_k}\right) \Delta \bar{\mu} & v_x(\mathbf{k}) > 0 \\
0 & v_x(\mathbf{k}) < 0
\end{cases}
\end{align}
where $v_x(\mathbf{k}) = \frac{1}{\hbar} \frac{\partial \epsilon_k}{\partial k_x}$.
Equipped with this solution, we are able to calculate the longitudinal and Hall response of our system. We have then:

\begin{align}
j_x &= -e \int \frac{d^2k}{(2 \pi)^2} v_x f_1 = -\frac{e}{h} \frac{\Delta \bar{\mu}}{2 \pi}  \int\limits_{v_x(\mathbf{k})>0} d^2 k \frac{\partial \epsilon_k}{\partial k_x} \left(-\frac{\partial f_0 }{\partial \epsilon_k}\right)
\end{align}

\begin{align}
j_y &= -e \int \frac{d^2k}{(2 \pi)^2} v_y f_1 = j_y^0 + j_H, \notag \\
\label{eq:j_trivial}
j_y^0 &= -\frac{e}{h} \frac{\Delta \bar{\mu}}{2 \pi} \int\limits_{v_x(\mathbf{k})>0} d^2 k \frac{\partial \epsilon_k}{\partial k_y}  \left(-\frac{\partial f_0 }{\partial \epsilon_k}\right), \\
\label{eq:j_magnus}
j_H &= \frac{e}{h} \frac{\Delta \bar{\mu}}{2 \pi} \int\limits_{v_x(\mathbf{k})>0} d^2 k \, \Omega(\mathbf{k})\frac{\partial U}{\partial s}  \left(-\frac{\partial f_0 }{\partial \epsilon_k}\right).
\end{align}
Here $j_y^0$ term arises from Fermi surface anisotropy and depends on its orientation relative to the direction of applied bias $\hat{x}$. Note that $j_y^0$ is independent of $\Delta U$. However, as we shall show later, this term can vanish due to symmetry.

To obtain the Hall current due to Berry curvature, we integrate the anomalous velocity contribution over the whole length of the device $I_y = \int_0^L dx j_H$ and then we can define Hall conductance as $G_H = -eI_y/\Delta\bar{\mu}$ and obtain at $T=0$:
\begin{equation}
\label{eq:hall_conductance}
G_H = \frac{e^2}{h} \frac{\Delta U}{2 \pi} \int\limits_{v_x(\mathbf{k})>0} d^2k \, \Omega(\mathbf{k})\, \delta\left( \epsilon_\mathbf{k} - \mu \right)
\end{equation}
where the electrostatic potential energy difference across the junction $\Delta U = U_S - U_D$ is assumed to be small.

This equation is the main result of this work. First of all, in the limit of small $\Delta U$ the Hall response does not depend on the detailed spatial dependence of the potential energy. Second, the dependence on $\Delta U$ is linear and the Hall current is independent of the system size in the ballistic limit. The effect is reduced by finite relaxation time as we show in Supplementary Materials. Crucially, Magnus Hall current is determined by directional Berry curvature of electrons with positive velocity along the direction of the applied bias. Therefore, by varying the bias with respect to crystal axis, Magnus Hall conductance can be used to %study the Fermi surface properties and
map Berry curvature distribution on the Fermi surface.

\begin{figure}
\includegraphics[width=0.4\textwidth]{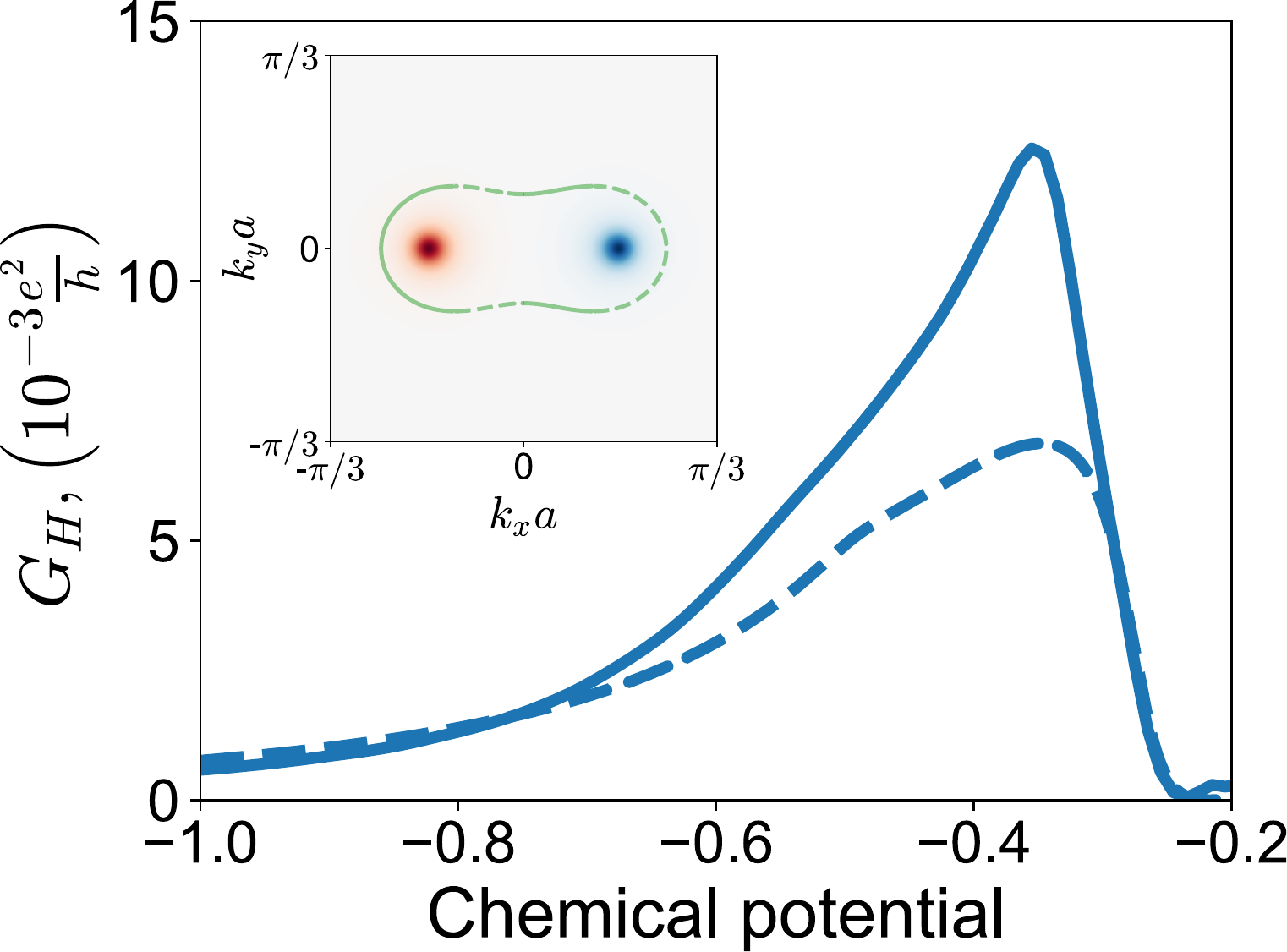}
\caption{\label{fig:transport_results}(Color online) Magnus Hall conductance $G_H$ as a function of the chemical potential $\mu$ at $T=0.01$ and $\Delta U = 0.05$ from Eq. \eqref{eq:hall_conductance} (dashed) and Landauer-Buttiker simulation (solid). Inset shows the distribution of Berry curvature of the valence bands in the Brillouin zone. Green line shows the Fermi surface for $E=-0.35$ (solid for $v_x>0$ and dashed for $v_x<0$).}
\end{figure}

\textit{Model}.--- To demonstrate this effect explicitly, we turn to a concrete model, which breaks inversion symmetry, but preserves time-reversal symmetry. We have chosen a simple two band model with the Hamiltonian:
\begin{align}
\label{eq:continuum_hamiltonian}
H(\mathbf{k}) = A k^2 + \left(B k^2 + \delta \right) \sigma_z + v_y k_y \sigma_y + D \sigma_x .
\end{align}
It contains two massive Dirac cones which are tilted when $A \neq 0$. This model captures essential features of the tilted Dirac cones of topological crystalline insulator surface states \cite{SodemannQuantumNonlinearHall2015} and low-energy band structure of 2D WTe$_2$ \cite{QianQuantumspinHall2014}. More details of the model are presented in the Supplemental Materials.

We can now compute Berry curvature $\mathbf{\Omega}(\mathbf{k})$ distribution in the Brillouin zone for parameters $A=0, B=1, \delta=-0.25, v_y=1.0, D=0.1$, which is shown in the inset of Fig.\ref{fig:transport_results} for the valence band. While the total Berry curvature integrated over Brillouin zone is equal to 0, the distribution consists of two peaks of opposite signs, located at the Dirac points.

Furthermore, this model possesses mirror symmetry $M_x: H(k_x, k_y) \rightarrow H(- k_x, k_y)$, which guarantees that $\epsilon(k_x, k_y) = \epsilon(-k_x, k_y)$. For $D=0$, this model has additional mirror symmetry $M_y: H(k_x, k_y) = \sigma_z H(k_x, -k_y) \sigma_z$. This symmetry is broken for $D\neq 0$. However, because the model is also time-reversal-invariant, we have $\epsilon(-k_x, k_y) = \epsilon(k_x, -k_y)$, which causes the current density $j_y^0$ of Eq.\eqref{eq:j_trivial} to vanish. Therefore, the Hall current will be determined solely by Magnus Hall contribution $j_H$ of Eq.\eqref{eq:j_magnus} and $G_H$ can be calculated according to Eq. \ref{eq:hall_conductance}. Result of such a calculation is presented as the dashed curve in Fig.\ref{fig:transport_results} as a function of chemical potential. Even though the total integral of $\Omega(\mathbf{k})$ vanishes, because our result for $G_H$ only relies on the Berry curvature of Bloch states with $v_x>0$, it is non-zero. For the set of parameters used in the calculation, both bands are symmetric with respect to $E=0$ line and so the result for the conduction band is mirror image of the curve for valence band. We note that the direction of the Hall current is the same for both bands, because while the Berry curvature switches sign to opposite between two bands, the velocities for given $\mathbf{k}$ also change to opposite and as a results the integration occurs over regions with the same values of $\Omega(\mathbf{k})$.

These approximate results derived from the semiclassical Boltzmann transport approach can be compared with a numerical tight-binding simulation using Landauer-Buttiker method. Details of the calculation are in Supplemental Materials. All the numerical simulations have been performed using Kwant package \cite{GrothKwantsoftwarepackage2014}. We consider the chemical potential dependence of $G^{LB}_H$, which we present as the solid curve in Fig. \ref{fig:transport_results} for $L=200$, $W=2400$ lattice sites and $\Delta U = 0.05$. The curve is obtained by temperature broadening with $T=0.01$. The qualitative behavior of the semiclassical result is reproduced with an asymmetric peak positioned away from the band bottom. The difference can be attributed to the differences in geometry of the setups for numerical simulation (finite width leads injecting current into the system) and semiclassical calculation (infinite width of the setup).

\begin{figure}
\includegraphics[width=0.49\textwidth]{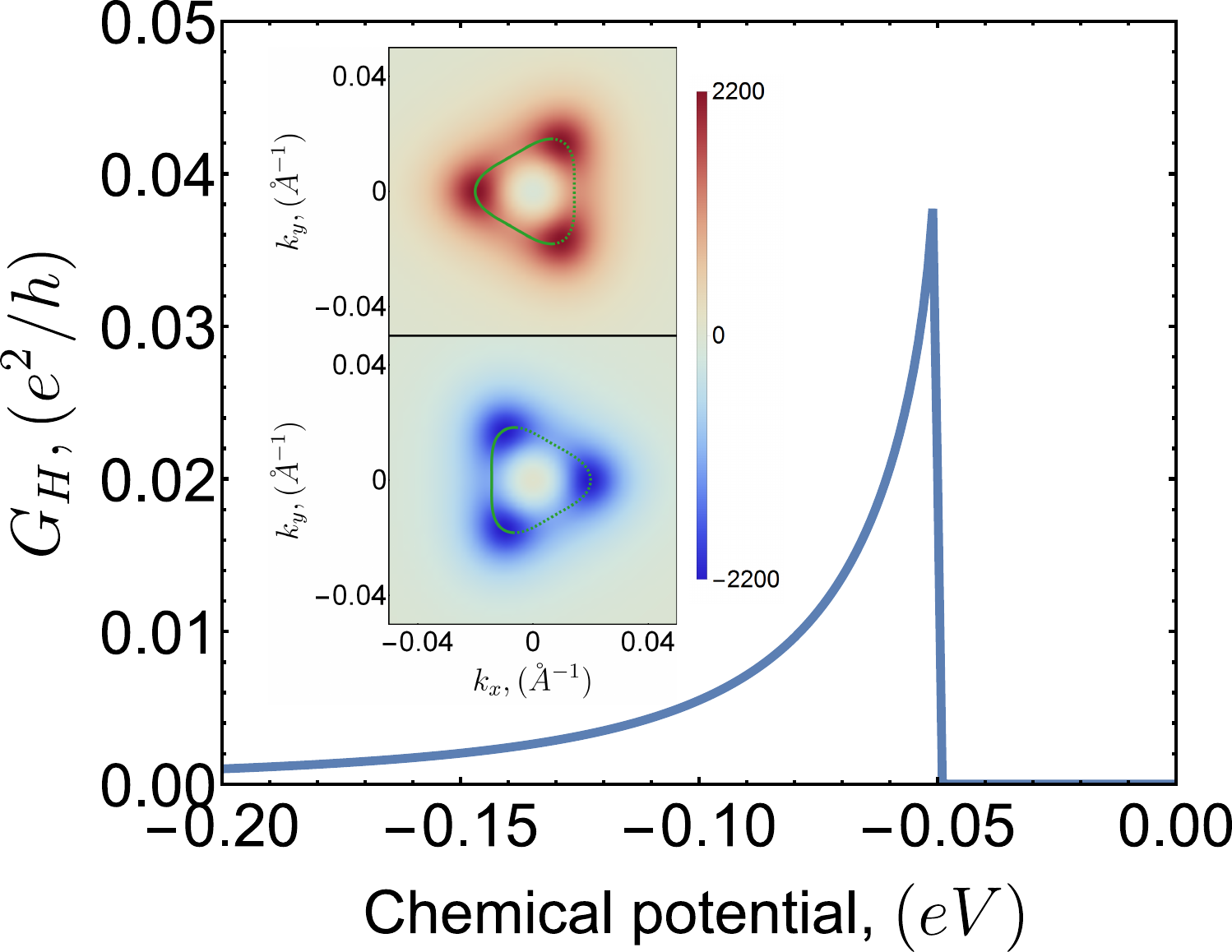}
\caption{\label{fig:graphene} Hall conductance for bilayer graphene model for $\Delta U = 10$ meV. Inset: Berry curvature distribution around $K$ and $K'$ valley of valence band of bilayer graphene. Green line shows the Fermi surface at $\mu=-60$ meV (solid for $v_x>0$ and dashed for $v_x<0$).}
\end{figure}

\textit{Candidate materials}.--- To observe the Magnus Hall effect under time-reversal symmetry, several conditions must be satisfied. First of all, the underlying material must break the inversion symmetry in order to have non-zero Berry curvature in the Brillouin zone. Furthermore, Berry curvature of right and left moving modes must be asymmetric. For example, under time-reversal symmetry massive Dirac fermions have to appear in pairs with opposite sign of Berry curvature. If they are isotropic, there will be an equal number of right movers with both signs of $\Omega(\mathbf{k})$ in each Dirac valley and their contributions will cancel each other. However, in general the Dirac cones are not perfectly isotropic and perfect cancellation will not occur. Two examples of materials that satisfy this condition are monolayer graphene on hBN (sublattice symmetry is broken due to formation of Moire superlatice) \cite{ParkNewGenerationMassless2008, YankowitzEmergencesuperlatticeDirac2012, HuntMassiveDiracFermions2013, YankowitzDynamicbandstructuretuning2018} and bilayer graphene with perpendicular electric field applied \cite{McCannLandauLevelDegeneracyQuantum2006, McCannelectronicpropertiesbilayer2013, RozhkovElectronicpropertiesgraphenebased2016}. In both cases trigonal warping introduces asymmetry between right and left movers in each valley. Moreover, these platforms support devices of high quality, which enable ballistic motion of electrons \cite{DuApproachingballistictransport2008, MayorovMicrometerScaleBallisticTransport2011, BanszerusBallisticTransportExceeding2016}, beneficial for observing the predicted effect.

As an example we use a model of bilayer graphene with trigonal warping and perpendicular electric field that opens up a gap. Calculations are performed using low-energy Hamiltonian that describes both $K$ and $K'$ valleys of bilayer graphene (labeled by $s=\pm 1$) \cite{McCannLandauLevelDegeneracyQuantum2006}:
\begin{equation}
H_s = \begin{pmatrix}
\Delta & sv k_{-s} - \lambda k_{s}^2\\
s v k_{s} - \lambda k_{-s}^2 & -\Delta
\end{pmatrix}
\end{equation}
where $k_\pm = k_x \pm i k_y$. Berry curvature distribution near both valleys is presented in the inset of Fig.\ref{fig:graphene}. The parameters used in the calculations are $\Delta=50$ meV, $v=10^5$ m/s and $\lambda = 1/(2m^*)$ with effective mass $m^*=0.033 m_e$, $m_e$ being electron mass \cite{Mucha-KruczynskiElectronholeasymmetry2010, McCannelectronicpropertiesbilayer2013}. The Fermi surface at $\mu=-60$ meV is indcated by the green line, solid for $v_x = \frac{1}{\hbar} \frac{\partial \epsilon_k}{\partial k_x} >0$ and dashed for $v_x<0$. We can now use this to compute the Hall conductance as a function of chemical potential in the vicinity of the band edge, which is shown in Fig. \ref{fig:graphene} for $\Delta U=10$ meV. Our result demonstrates explicitly that Magnus Hall effect does not rely on Berry curvature dipole \cite{SodemannQuantumNonlinearHall2015, FacioStronglyEnhancedBerry2018, TsirkinGyrotropiceffectstrigonal2018, YouBerrycurvaturedipole2018, ZhangBerrycurvaturedipole2018, ZhangElectricallytuneablenonlinear2018, ShiSymmetryspintexturetunable2019} or the presence of skew scattering \cite{IsobeHighfrequencyrectificationchiral2018, NandySymmetryQuantumKinetics2019, KonigGyrotropicHalleffect2018, DuDisorderinducednonlinearHall2018} that are necessary conditions for nonlinear Hall effect.

\textit{Summary}.--- In this paper we have demonstrated existence of Magnus Hall effect in inversion symmetry breaking, but time reversal invariant systems that have non-zero Berry curvature. The effect relies on built-in electric field in the device and should be most pronounced in ballistic systems. The significance of the Magnus Hall effect is twofold. It opens a pathway to a new generation of current rectification devices. It also provides a much needed tool to map Berry curvature distribution of quantum materials in momentum space.

\textit{Acknowledgments}.--- We thank Hiroki Isobe and Su-Yang Xu for helpful discussions. This work was supported by DOE Office of Basic Energy Sciences, Division of Materials Sciences and Engineering under Award DE-SC0018945. MP was supported by Shell through the MIT Energy Initiative. LF was supported in part by a Simons Investigator Award from the Simons Foundation.

\bibliography{nonlinear_Hall}

\newpage
\setcounter{equation}{0}
\setcounter{figure}{0}
\setcounter{page}{1}
\renewcommand{\thefigure}{S\arabic{figure}}
\renewcommand{\theequation}{S\arabic{equation}}

\begin{center}
\textbf{\large Supplemental Material for \\ "Magnus Hall effect"}
\end{center}

\textit{Finite relaxation time}.---
To take into account the effects of scattering, we use Boltzmann equation (BE) in relaxation time approximation:
\begin{equation}
\label{eq:SM_boltzmann_eq}
\frac{\partial f}{\partial t} + \dot{\mathbf{r}} \cdot \frac{\partial f}{\partial \mathbf{r}} + \dot{\mathbf{k}} \cdot \frac{\partial f}{\partial \mathbf{k}} = - \frac{f-f_0}{\tau}
\end{equation}
where $f(\mathbf{k}, \mathbf{r})$ is the occupation distribution function, $f_0$ is the equilibrium distribution and $\tau$ is the relaxation time. We solve the Boltzmann equation in the same geometry as in the main text, but allow for a rotation of the direction of applied bias $\Delta \bar{\mu}$ (parallel to the direction of built-in electric field) with respect to Fermi surface orientation. We label the direction parallel to the bias direction as $\hat{n} = \cos(\phi)\hat{x} + \sin(\phi) \hat{y}$, where $\phi$ is an arbitrary angle that represents rotation of the device with respect to underlying crystal, which we keep fixed.  This allows to account for dependence of the Hall current on anisotropy of the Fermi surface. The gate-induced electrostatic potential is now labeled $U(s)$ (with $s$ being the coordinate along $\hat{n}$, so that $-\nabla U = F(s) \hat{n}$). Taking into consideration the device geometry of our system, the Boltzmann equation assumes the form:

\begin{equation}
\frac{1}{\hbar} \frac{\partial f}{\partial s} \frac{\partial \epsilon_k}{\partial \mathbf{k}}  \cdot \hat{n}  + \frac{1}{\hbar} F(s) \frac{\partial f}{\partial \mathbf{k}} \cdot \hat{n} = - \frac{f - f_0}{\tau}
\end{equation}

We again look for a solution in the lowest order of perturbation in $\Delta \bar{\mu}$, so we write $f = f_0 + f_1$, where $f_1$ is the nonequilibrium part first order in the perturbation. This gives an equation for $f_1$:

\begin{equation}
\frac{1}{\hbar} \frac{\partial f_1}{\partial s} \frac{\partial \epsilon_k}{\partial \mathbf{k}}  \cdot \hat{n}  + \frac{1}{\hbar} F(s) \frac{\partial f_1}{\partial \mathbf{k}} \cdot \hat{n} = - \frac{f_1}{\tau}
\end{equation}

Due to weak variation of $U(s)$, we can drop second term on the right hand side and using $v_n = \frac{1}{\hbar} \frac{\partial \epsilon_k}{\partial \mathbf{k}}  \cdot \hat{n}$ we arrive at:

\begin{equation}
\frac{\partial f_1}{\partial s} = - \frac{f_1}{v_n \tau}
\end{equation}

The general solution of this equation is:
\begin{equation}
f_1(\mathbf{k}, \mathbf{r}) = A(\mathbf{k}) e^{-\frac{s}{v_n \tau}} + B(\mathbf{k})
\end{equation}

We can now determine $A$ and $B$ from the boundary condition similarly to the main text. Since the source has a larger electrochemical potential, it results in a surplus of electrons entering the system at $s=0$ with positive $v_n$ velocity. If the system is ballistic ($\tau \rightarrow \infty$), these electrons propagate across the device without any scattering. When disorder is introduced into the sample, finite relaxation time will result in exponential decay of the surplus distribution. Therefore we have:

\begin{equation}
f_1(\mathbf{k}, \mathbf{r}) = 
\begin{cases}
\left(-\frac{\partial f_0 }{\partial \epsilon_k}\right) \Delta \bar{\mu} \, e^{-\frac{s}{v_n \tau}} & v_n(\mathbf{k}) > 0 \\
0 & v_n(\mathbf{k}) < 0
\end{cases} 
\end{equation}

Therefore, in the simplest generalization to the case with finite relaxation time the distribution of surplus electrons that exit the source region will decay exponentially on the length scale comparable to mean free path $l_{m} \approx v_F \tau$. Thus, the effect of disorder on Magnus Hall effect will be suppressed as long as the device length is shorter than mean free path.

\begin{figure}
\includegraphics[width=0.4\textwidth]{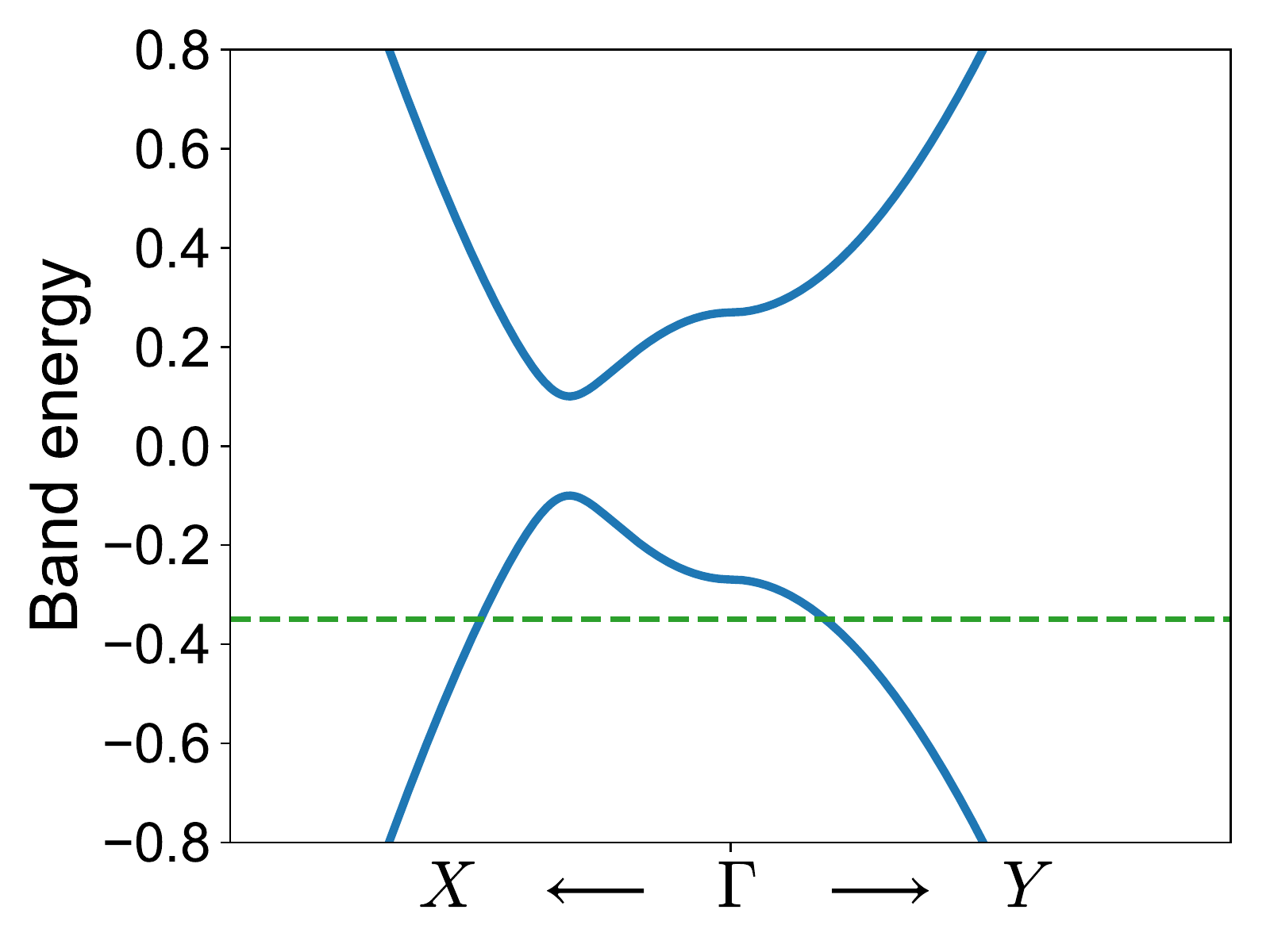}
\caption{\label{fig:bands} Band structure of model \eqref{eq:continuum_hamiltonian} along $\Gamma - X$ and $\Gamma - Y$ directions for parameters $A=0, B=1, \delta=-0.25, v_y=1.0, D=0.1$. Dashed line indicates $\mu=-0.35$ for which Fermi surface contour was plotted on Berry curvature distribution in the main text.}
\end{figure}

\textit{Additional details of the two band model}.---As an example of a model which exhibits Magnus Hall effect we proposed a model with the Hamiltonian:
\begin{align}
\label{eq:SM_continuum_hamiltonian}
H(\mathbf{k}) = A k^2 + \left(B k^2 + \delta \right) \sigma_z + v_y k_y \sigma_y + D \sigma_x
\end{align}
which preserves time-reversal, but breaks inversion symmetry for $D \neq 0$. It consists of 2 bands of different effective masses determined by $A$ and $B$ which hybridize with a $k_y$-dependent term. Parameter $\delta$ controls the band gap at $\Gamma$ point and if $\delta<0$ the bands are inverted around that point. When $D=0$ the spectrum of this Hamiltonian is degenerate at two points in the Brillouin zone and the dispersion can be linearized in their vicinity, giving two Dirac cones. These cones are tilted when $A \neq 0$. To get non-zero Berry curvature, one has to break inversion symmetry to observe the effects described in the main text section. Since we are interested in the Hall effect while time-reversal symmetry is preserved and the integral of Berry curvature over Brillouin zone vanishes, we have to break inversion symmetry with $D \neq 0$. This adds mass to both of the Dirac cones, opening a band gap. This is presented in Fig.\ref{fig:bands}, which shows the band structure along $\Gamma - X$ and $\Gamma - Y$ directions for parameters $A=0, B=1, \delta=-0.25, v_y=1.0, D=0.1$. This choice of parameters corresponds to rotationally symmetric spectrum close to the Dirac points.

\begin{figure}
\includegraphics[width=0.45\textwidth]{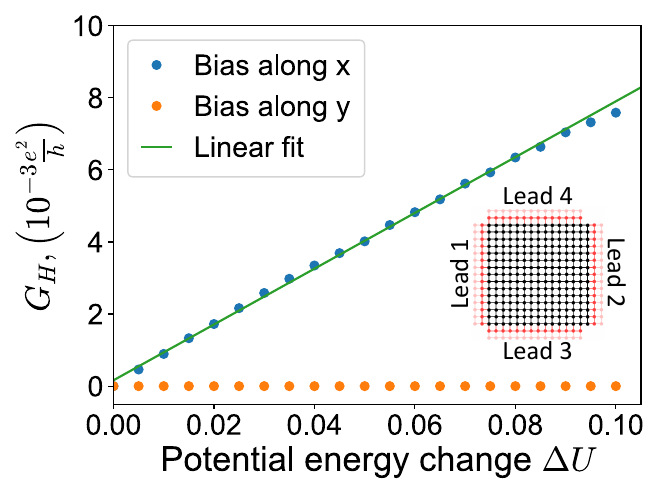}
\caption{\label{fig:potential_energy_change}(Color online) Hall conductance obtained using Landauer-Buttiker numerical simulation. Dependence on potential energy change $\Delta U$ for chemical potential bias applied along $x$ and $y$ axis. Solid line is a fit of linear dependence.}
\end{figure}

\textit{Details of numerical simulations and additional results}.---
To perform Landauer-Buttiker simulation, we discretize the continuum Hamiltonian \eqref{eq:continuum_hamiltonian} on a square lattice with lattice constant $a=1$. The potential energy drop is included as a spatially varying onsite energy given by:
\begin{equation}
U(x) = -\Delta U \tanh\left(\frac{x/L-x_c}{s}\right)
\end{equation}
In the simulations we used $x_c = 0.5$ and $s=0.1$. The scattering region is a rectangle that consists of $L \times W$ lattice sites and to each of the sides of the device we attach a semi-infinite lead of width that is equal to the length of that particular side (as shown in inset of Fig. \ref{fig:potential_energy_change}). The leads are described by the same Hamiltonian as the scattering region and they also include $U(x)$ potential to ensure no mismatch at the lead-scattering region interface. In such a setup we compute the scattering matrices for varying chemical potential $\mu$, system sizes and potential energy change $\Delta U$. From scattering matrices we obtain transmission probabilities $T_{ij}$ for each pair of leads and then calculate $T^{-1}$. We use it to calculate voltages at each of the leads $V_i$ in two cases: (1) current $I_x$ flowing between leads 1 and 2 (leads 3 and 4 serve as voltage probes only) and (2) current $I_y$ flowing between leads 3 and 4 (leads 1 and 2 serve as voltage probes only). For each case we then define $\Delta V^{(k)}_{12} = V^{(k)}_{2} - V^{(k)}_{1}$ and $\Delta V^{(k)}_{34} = V^{(k)}_{4} - V^{(k)}_{3}$. We can now setup a set of linear equations:
\begin{equation}
\label{eq:conductances}
\begin{aligned}
I_x = G_{xx} \Delta V^{(1)}_{12} + G_{xy} \Delta V^{(1)}_{34} \\
0 = G_{yx} \Delta V^{(1)}_{12} + G_{yy} \Delta V^{(1)}_{34} \\
0 = G_{xx} \Delta V^{(2)}_{12} + G_{xy} \Delta V^{(2)}_{34} \\
I_y = G_{yx} \Delta V^{(2)}_{12} + G_{yy} \Delta V^{(2)}_{34}
\end{aligned}
\end{equation}
From this we can then obtain $G_{xx}$, $G_{xy}$, $G_{yx}$ and $G_{yy}$ for the sample. The value $G_{yx}$ will correspond to $G_H$ as used for comparison with semiclassical calculation in the main text.

In Fig. \ref{fig:potential_energy_change} we show dependence of Hall conductance on the potential energy change $\Delta U$ for $\mu = -0.6$ and system size $L=200$, $W=2400$. When chemical potential imbalance is applied along $x$ direction, Hall conductance increases linearly with $\Delta U$ before eventually acquiring some nonlinear contribution. This is shown by a least-squares linear fit to the simulation results and is consistent with the semiclassical calculation. Finally, when we rotate our setup by 90$^\circ$, effectively applying bias and potential difference along $y$ direction of the model, the result is zero independent of potential energy change, which in semiclassical formalism corresponds to vanishing of the Berry curvature integral due to symmetry.

\begin{figure}
\includegraphics[width=0.45\textwidth]{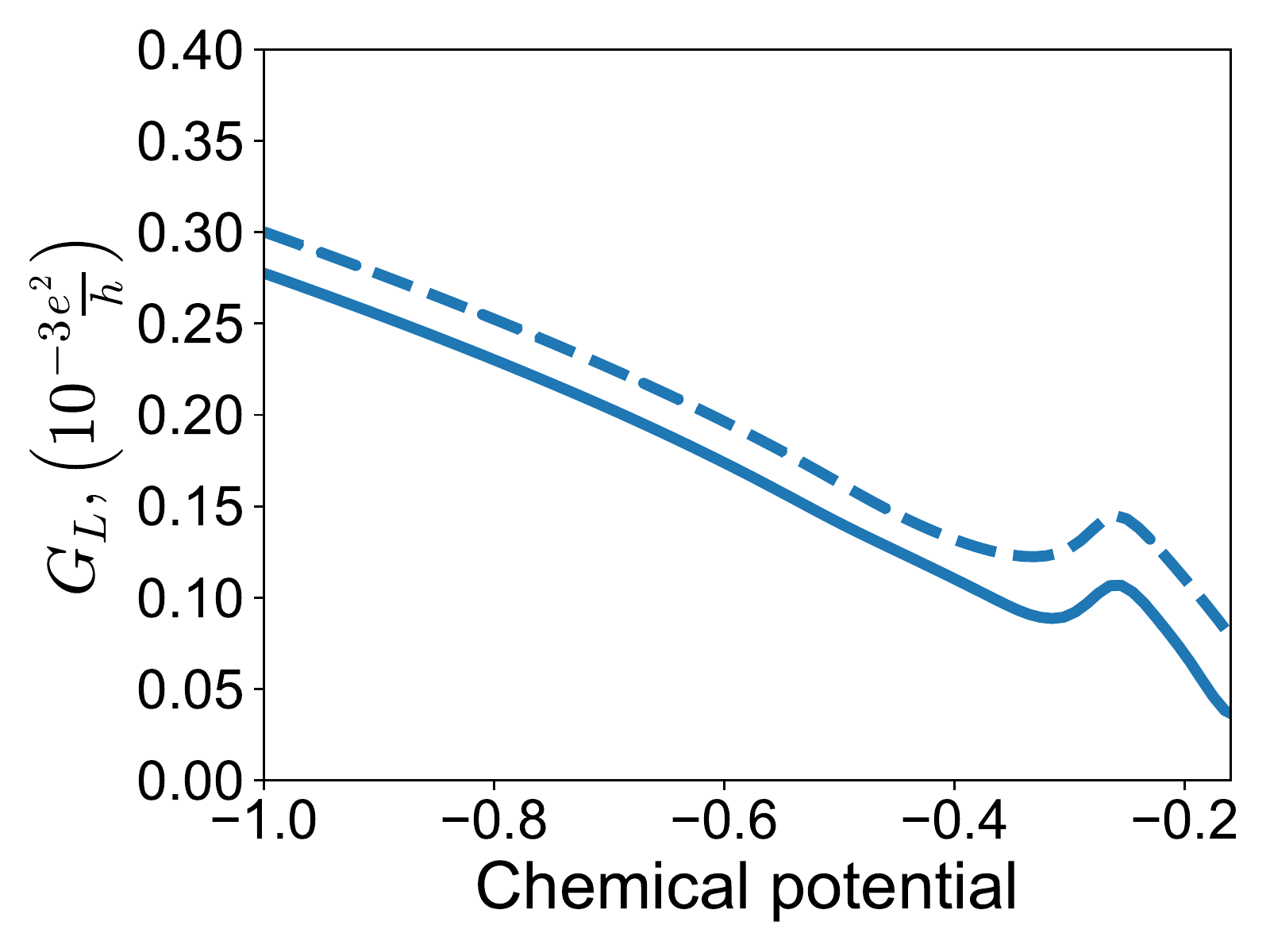}
\caption{\label{fig:long_conductance}(Color online) Longitudinal conductance obtained using Landauer-Buttiker numerical simulation (solid line) and from semiclassical calculation (dashed line) as a function of the chemical potential $\mu$ at $\Delta U = 0.05$.}
\end{figure}

In Fig.\ref{fig:long_conductance} we present longitudinal conductance $G_L = G_{xx}/W$ obtained using the numerical simulation as a solution of Eq.\eqref{eq:conductances} (solid line) and from semiclassical calculation presented in the main text (dashed line). The deviation is mainly due to omission of the built-in electric field effect in the semiclassical calculaton.

\end{document}